# Force Density Balance inside the Hydrogen Atom

F. J. Himpsel

Department of Physics, University of Wisconsin Madison,
1150 University Ave., Madison, WI 53706, USA, fhimpsel@wisc.edu

**Abstract**

Motivated by the long-debated question about the internal stability of the electron, the force densities acting on the charge density of the $1s$ electron in the H atom are investigated. The problem is mapped onto the canonical formalism for a classical Dirac field coupled to the electric field of an external point charge. An explicit calculation shows that the attractive Coulomb force density is balanced exactly at every point in space by the repulsive confinement force density. The latter requires evaluating the divergence of the stress tensor for the $1s$ solution of the Dirac equation. Such a local force balance goes beyond the global stability criteria that are usually given for the H atom. This concept is extended to the internal stability of any charged particle by investigating the force densities acting on its surrounding vacuum polarization. At large distances one has to consider only the charge density of virtual electrons and positrons, induced by a point charge in the vacuum of quantum electrodynamics.

## 1. Motivation

The internal stability of the electron has been a long-standing mystery [1],[2],[3]. Viewed as a negative charge cloud, it would Coulomb-explode due to electrostatic repulsion. The repulsive forces become infinitely large when the radius r of the charge cloud shrinks to zero. More generally, any charged elementary particle would be susceptible to such an instability. The problem is often cast in terms of the self-energy of an electron and the resulting mass: The electromagnetic energy of a charge cloud diverges like $r^{-1}$ in classical electrodynamics. That would imply an infinite mass for a point charge. Quantum electrodynamics [3] reduces this divergence from $r^{-1}$ to $\ln(r)$ and then eliminates it altogether by mass renormalization. But such a manipulation bypasses the question whether or not an internal force density balance exists for the electron. In fact, the electron is not a pure point charge in quantum electrodynamics. It surrounds itself with a faint halo of vacuum polarization that extends out to distances comparable to the reduced Compton wavelength $\lambdabar_C = \hbar/m_e c \approx 4 \cdot 10^{-13}$ m.

Before getting into such a complex manybody problem it is worthwhile to investigate a simple, single-particle system to find out whether it makes sense talking about the internal stability of an electron. The 1s electron in the H atom serves this purpose well. In that case the Coulomb energy is attractive and threatens a collapse. Explaining the stability of the H atom played an important role in establishing quantum mechanics. Today we would say that the H atom is stable, because it represents a minimum of the total energy [4]. The attractive Coulomb energy is balanced by the repulsive kinetic energy that is generated by confining the 1s electron to the Bohr radius $a_0 = \lambdabar_C/\alpha$ ($\alpha = e^2/\hbar c \approx 1/137$). Such reasoning does not reveal anything about the local stability of the 1s charge density, though. This question tends to be brushed off as unphysical, since the uncertainty relation seems to forbid looking inside the H atom without disturbing it.

On the other hand, it is possible to define a local force density acting on the Dirac field $\psi$ which describes the 1s electron in the single-particle picture. It has again two opposing parts: 1) The attractive Coulomb force density $\rho \cdot \mathbf{E}$ exerted by the external electric field $\mathbf{E}$ onto the charge density $\rho = -e \psi^* \psi$. 2) A repulsive confinement force density exerted by Dirac field $\psi$ onto itself. The latter is more difficult to calculate, but it is well-defined as the divergence of the stress tensor. That in turn can be obtained via the canonical formalism. It is found that the two force densities indeed match each other exactly in the H atom.

## 2. Force Density Balance

To keep the expressions transparent, the proton is modeled by a fixed, positive point charge without a magnetic moment. The Dirac wave function $\psi$ of the H 1s electron is treated as classical Dirac field which interacts with the Coulomb field $\mathbf{E}$ of the external point charge. This makes it possible to use the canonical formalism of relativistic field theory [5] for defining the force densities associated with $\psi$ and $\mathbf{E}$. These are obtained by constructing the energy-momentum tensor $T^{\mu\nu}$ from the Lagrangian and taking the divergence of its spatial part (the stress tensor $-T^{ij}$). The symmetric, gauge-invariant

Lagrangian of the Dirac field $\psi$ interacting with an external electromagnetic potential $A_\mu$ has the following form [5]:

(1) $\quad L_D = \tfrac{1}{2}[\bar{\psi}\gamma^\mu\cdot(iD_\mu)\psi + (iD_\mu)^*\bar{\psi}\cdot\gamma^\mu\psi] - m_e\bar{\psi}\psi \qquad\qquad iD_\mu = i\partial_\mu - qA_\mu$

$\qquad\quad = \{\tfrac{1}{2}i[\bar{\psi}\gamma^\mu\cdot(\partial_\mu\psi) - (\partial_\mu\bar{\psi})\cdot\gamma^\mu\psi] - m_e\bar{\psi}\psi\} - q(\bar{\psi}\gamma^\mu\psi)\cdot A_\mu \qquad q = -e \text{ for } e^-$

$\qquad\quad = L_\psi + L_{A\psi}$

The differentiation is not carried past the next multiplication sign. The Lagrangian $L_D$ consists of the pure Dirac term $L_\psi$ and the interaction term $L_{A\psi}$. The Lagrangian for the external electromagnetic field $A_\mu$ is not included.

The Euler-Lagrange equation of $L_D$ is the Dirac equation in the presence of an electromagnetic potential $A_\mu$. Together with the conjugate equation for $\bar{\psi}$ one obtains two additional relations after multiplication with $\bar{\psi}$ and $\psi$, respectively. The sum shows that $L_D$ vanishes, and the difference gives the continuity equation for the Dirac current density $j^\nu$:

(2) $\quad \partial L_D/\partial\bar{\psi} - \partial_\mu[\partial L_D/\partial(\partial_\mu\bar{\psi})] = 0$

$\qquad \partial L_D/\partial\psi - \partial_\mu[\partial L_D/\partial(\partial_\mu\psi)] = 0$

$\qquad \left.\begin{array}{l}(+i\partial_\mu - qA_\mu)\gamma^\mu\psi - m_e\psi = 0 \\ (-i\partial_\mu - qA_\mu)\bar{\psi}\gamma^\mu - m_e\bar{\psi} = 0\end{array}\right\} \Rightarrow \quad L_D = 0 \qquad \partial_\nu j^\nu = 0$

(3) $\quad j^\mu = -\partial L/\partial A_\mu = q(\bar{\psi}\gamma^\mu\psi) = (\rho, \mathbf{J}) \qquad\qquad A_\nu = (\Phi, -\mathbf{A})$

The symmetric energy-momentum tensor $T_D^{\mu\nu}$ corresponding to $L_D$ is obtained via the canonical formalism [5]:

(4) $\quad T_D^{\mu\nu} = \tfrac{1}{4}\{[\bar{\psi}\gamma^\mu\cdot(iD^\nu)\psi + \bar{\psi}\gamma^\nu\cdot(iD^\mu)\psi] + [(iD^\mu)^*\bar{\psi}\cdot\gamma^\nu\psi + (iD^\nu)^*\bar{\psi}\cdot\gamma^\mu\psi]\}$

$\qquad\quad = \tfrac{1}{4}i\cdot\{[\bar{\psi}\gamma^\mu\cdot(\partial^\nu\psi) + \bar{\psi}\gamma^\nu\cdot(\partial^\mu\psi)] - [(\partial^\mu\bar{\psi})\cdot\gamma^\nu\psi + (\partial^\nu\bar{\psi})\cdot\gamma^\mu\psi]\} - \tfrac{1}{2}(j^\mu A^\nu + j^\nu A^\mu)$

$\qquad\quad = T_\psi^{\mu\nu} + T_{A\psi}^{\mu\nu}$

The exclusive use of the covariant derivative $D_\mu$ makes this energy-momentum tensor explicitly gauge-invariant. Like the Lagrangian, it separates into a pure Dirac term $T_\psi^{\mu\nu}$ and an interaction term $T_{A\psi}^{\mu\nu}$ which contains the external four-potential $A_\mu$. The canonical term $g^{\mu\nu}L_D$ vanishes, since $L_D = 0$ in (2). Noether's theorem then leads to the following continuity equation for the energy-momentum tensor [5]:

(5) $\quad \partial_\mu T_D^{\mu\nu} = \partial_\mu T_\psi^{\mu\nu} + j_\mu F^{\mu\nu} = 0$

The decomposition of $T_\psi^{\mu\nu}$ into temporal and spatial parts yields the energy density $H_\psi$, the momentum density $\mathbf{S}_\psi$, and the stress tensor $\mathbf{T}_\psi = -T_\psi^{ij}$ of the Dirac field $\psi$. The latter has been defined with a minus sign analogous to the Maxwell stress tensor. The divergence of the stress tensor then defines the confinement force density $\mathbf{f}_\psi$ exerted by $\psi$ onto itself. It is opposed by the Lorentz force density $\mathbf{f}_{A\psi}$ exerted by the external electromagnetic field $F^{\mu\nu}$ onto $\psi$. This is the force density balance that will guarantee the local stability of the H 1s electron.

(6) $\quad T_\psi^{\mu\nu} = \begin{bmatrix} H_\psi & S_\psi \\ S_\psi & -T_\psi \end{bmatrix} \qquad\qquad F^{\mu\nu}=\partial^\mu A^\nu-\partial^\nu A^\mu \quad \mathbf{E}=-\nabla\Phi-\partial\mathbf{A}/\partial t \quad \mathbf{B}=\nabla\times\mathbf{A}$

$\qquad -\partial_i T_\psi^{ik} = \quad \nabla\cdot\mathbf{T}_\psi \quad = \mathbf{f}_\psi \qquad\qquad$ Confinement force density

$\qquad -j_\mu F^{\mu k} = (\rho\mathbf{E}+\mathbf{J}\times\mathbf{B}) = \mathbf{f}_{A\psi} \qquad$ Lorentz force density

The temporal and spatial parts of $T_\psi^{\mu\nu}$ and their continuity equations can be written as:

(7) $\quad H_\psi = \tfrac{1}{2}i\,[\psi^*\cdot(\partial\psi/\partial t) - (\partial\psi^*/\partial t)\cdot\psi]$

$\qquad \mathbf{S}_\psi = \tfrac{1}{4}i\,\{[\overline{\psi}\boldsymbol{\gamma}\cdot(\partial\psi/\partial t) - (\partial\overline{\psi}/\partial t)\cdot\boldsymbol{\gamma}\psi] - [\psi^*\cdot(\nabla\psi)-(\nabla\psi^*)\cdot\psi]\}$

$\qquad T_\psi^{ij} = \tfrac{1}{4}i\,\{[\overline{\psi}\gamma^i\cdot(\nabla_j\psi) - (\nabla_i\overline{\psi})\cdot\gamma^j\psi] + [\overline{\psi}\gamma^j\cdot(\nabla_i\psi) - (\nabla_j\overline{\psi})\cdot\gamma^i\psi]\}$

$\qquad 0 = \partial H_\psi/\partial t + \nabla\cdot\mathbf{S}_\psi + \mathbf{J}\cdot\mathbf{E}$

$\qquad 0 = \partial\mathbf{S}_\psi/\partial t - \nabla\cdot\mathbf{T}_\psi - (\rho\mathbf{E}+\mathbf{J}\times\mathbf{B})$

A change of the energy density $\partial H_\psi/\partial t$ is converted into a source of momentum density $\nabla\cdot\mathbf{S}_\psi$ plus an electric power density $\mathbf{J}\cdot\mathbf{E}$. The change of the momentum density $\partial\mathbf{S}_\psi/\partial t$ is equal to the two force densities $\mathbf{f}_\psi=\nabla\cdot\mathbf{T}_\psi$ and $\mathbf{f}_{A\psi}=(\rho\mathbf{E}+\mathbf{J}\times\mathbf{B})$, representing a local version of Newton's second law.

Consider now a stationary Dirac field $\psi$ with the time dependence $\exp(-iEt)$, where E is the energy eigenvalue of $\psi$. $T_\psi^{\mu\nu}$ becomes time-independent, and the two continuity equations in (7) represent energy conservation and the force density balance:

(8) $\quad H_\psi = E\cdot\psi^*\psi$

$\qquad \mathbf{S}_\psi = \tfrac{1}{2}E\cdot\overline{\psi}\boldsymbol{\gamma}\psi - \tfrac{1}{4}i[\psi^*\cdot(\nabla\psi)-(\nabla\psi^*)\cdot\psi]$

$\qquad T_\psi^{ij}$ remains the same as in (6), but it is now independent of t.

(9) $\quad 0 = \nabla\cdot\mathbf{S}_\psi + \mathbf{J}\cdot\mathbf{E} \qquad\qquad$ Energy conservation

$\qquad 0 = \nabla\cdot\mathbf{T}_\psi + (\rho\mathbf{E}+\mathbf{J}\times\mathbf{B}) = \mathbf{f}_\psi + \mathbf{f}_{A\psi} \qquad$ Force density balance

In order to test the force density balance explicitly, $\mathbf{f}_\psi$ and $\mathbf{f}_{A\psi}$ are evaluated for the 1s solution of the Dirac equation in the presence of the Coulomb potential $\Phi$ of a point charge $+e$:

(10) $\quad \Phi = e\,r^{-1} \qquad q=-e \qquad q\Phi = -\alpha\,r^{-1} \qquad r=|\mathbf{r}|$

In the absence of a magnetic field the Lagrangian and the Dirac equation take the form:

(11) $\quad L_D = \overline{\psi}\gamma^0(E+\alpha r^{-1})\psi + \tfrac{1}{2}(\overline{\psi}\boldsymbol{\gamma}\cdot i\nabla\psi - i\nabla\overline{\psi}\cdot\boldsymbol{\gamma}\psi) - m_e\overline{\psi}\psi$

$\qquad \{\gamma^0(E+\alpha/r) + i\boldsymbol{\gamma}\cdot\nabla - m_e\}\psi = 0 \qquad \boldsymbol{\gamma}\cdot\nabla = \gamma^r\partial_r + r^{-1}\cdot\gamma^\theta\partial_\theta + (r\sin\theta)^{-1}\cdot\gamma^\varphi\partial_\varphi$

A conversion to spherical coordinates involves the Dirac matrices $\gamma^r,\gamma^\theta,\gamma^\varphi$ defined in the Appendix. The solutions of the Dirac equation are obtained by separating radial and angular variables via the the angular momentum operator **L** and its relativistic analog K. The eigenvalues $-\kappa$ of K determine the angular momentum quantum numbers $l$ and $j$:

(12) $\quad \mathbf{L} = \mathbf{r} \times (-i\nabla) \qquad \nabla = \mathbf{e}_r \cdot \partial_r - i r^{-1} \cdot \mathbf{e}_r \times \mathbf{L} \qquad \gamma \cdot i\nabla = i\gamma^r(\partial_r - r^{-1}\boldsymbol{\sigma} \cdot \mathbf{L})$

$\qquad K = \gamma^0(\boldsymbol{\sigma} \cdot \mathbf{L} + 1) \qquad K\psi = -\kappa\psi \qquad l = |\kappa + \tfrac{1}{2}| - \tfrac{1}{2} \qquad j = |\kappa| - \tfrac{1}{2} \qquad m_j = m$

$\qquad \{\gamma^0(E + \alpha/r) + i\gamma^r \cdot [\partial_r + (1+\gamma^0\kappa)/r] - m_e\}\psi = 0$

Writing $\psi(r,\theta,\varphi)$ as a product of the radial functions $g(r), f(r)$ with the spherical Pauli spinors $\chi_\kappa^m(\theta,\varphi), \chi_{-\kappa}^m(\theta,\varphi)$ from the Appendix yields the radial Dirac equation:

(13) $\quad \psi(r,\theta,\varphi) = \{g(r) \cdot \chi_\kappa^m(\theta,\varphi), i f(r) \cdot \chi_{-\kappa}^m(\theta,\varphi)\}$

(14) $\quad \partial_r \begin{bmatrix} rf \\ rg \end{bmatrix} = \begin{bmatrix} +\frac{\kappa}{r} & m_e - (E + \frac{\alpha}{r}) \\ m_e + (E + \frac{\alpha}{r}) & -\frac{\kappa}{r} \end{bmatrix} \cdot \begin{bmatrix} rf \\ rg \end{bmatrix}$

The 1$s$ state ($\kappa = -1, m = \tfrac{1}{2}$) has the following wave functions and charge/current densities:

(15) $\quad \gamma_{1s} = (1-\alpha^2)^{1/2} \qquad E_{1s} = \gamma_{1s} \cdot m_e \qquad p_{1s} = (m_e^2 - E_{1s}^2)^{1/2} = \alpha \cdot m_e \qquad A_{1s} = [\alpha/\Gamma(1+2\gamma_{1s})]^{1/2}$

$\qquad r\, g_{1s}(r) = +A_{1s}(m_e + E_{1s})^{1/2} \cdot (2p_{1s}r)^\gamma \exp(-p_{1s}r) \qquad \chi_{-1}^{1/2}(\theta,\varphi) = (4\pi)^{-1/2} \cdot \{1, 0\}$

$\qquad r\, f_{1s}(r) = -A_{1s}(m_e - E_{1s})^{1/2} \cdot (2p_{1s}r)^\gamma \exp(-p_{1s}r) \qquad \chi_{+1}^{1/2}(\theta,\varphi) = (4\pi)^{-1/2}\{-\cos\theta, -\sin\theta\, e^{i\varphi}\}$

(16) $\quad \rho_{1s} = -e(\psi_{1s}^*\psi_{1s}) \;= -e\frac{1}{4\pi}(f_{1s}^2 + g_{1s}^2) \qquad \int\rho_{1s}d^3r = -e\int(f_{1s}^2 + g_{1s}^2) \cdot r^2 dr = -e$

(17) $\quad J_{1s}^\varphi = -e(\bar\psi_{1s}\gamma^\varphi\psi_{1s}) = +e\frac{1}{2\pi} f_{1s} g_{1s} \sin(\theta) \qquad J_{1s}^r = 0 \qquad J_{1s}^\theta = 0$

Since $f_{1s}$ and $g_{1s}$ have opposite signs, the azimuthal current density $J_{1s}^\varphi$ acquires the correct negative sign for $m = +\tfrac{1}{2}$, taking the negative charge of the electron into account. Despite $l = 0$ the azimuthal current density $J_{1s}^\varphi$ does not vanish, since spin-orbit coupling mixes the upper Pauli spinor ($\kappa = -1, l = 0$) with the lower one ($\kappa = +1, l = 1$).

The stress tensor $\mathbf{T}_\psi$ in (6) contains the tensor product of the vector $\{\gamma^r, \gamma^\theta, \gamma^\varphi\}$ with the gradient operator $\{\partial_r, r^{-1}\partial_\theta, (r\sin\theta)^{-1}\partial_\varphi\}$, applied to $\psi$. For $s$-states, $\mathbf{T}_\psi$ has only diagonal components. The confinement force density $\mathbf{f}_\psi$ in (6) is given by the tensor divergence of $\mathbf{T}_\psi$ which takes the following form for a diagonal tensor:

$\nabla \cdot \mathbf{T}_\psi = \{r^{-2}\partial_r(r^2 T_\psi^{rr}) - r^{-1}(T_\psi^{\theta\theta} + T_\psi^{\varphi\varphi}),\; (r\sin\theta)^{-1}\partial_\theta(\sin\theta\, T_\psi^{\theta\theta}) - (r\tan\theta)^{-1}T_\psi^{\varphi\varphi},\; (r\sin\theta)^{-1}\partial_\varphi T_\psi^{\varphi\varphi}\}$

(18) $\quad T_\psi^{rr} = -\frac{1}{4\pi}(f \cdot g' - f' \cdot g)$

$\qquad T_\psi^{\theta\theta} = +\frac{1}{4\pi r} f \cdot g$

$\qquad T_\psi^{\varphi\varphi} = +\frac{1}{4\pi r} f \cdot g$

(19) $\quad f_\psi^r = -\frac{1}{4\pi}[2fg/r^2 + 2(fg' - f'g)/r + (fg'' - f''g)]$

$\qquad f_\psi^\theta = 0$

$\qquad f_\psi^\varphi = 0$

The Lorentz force density reduces to the radial Coulomb force density $\rho_{1s} \cdot \mathbf{E}$, with $\rho_{1s}$ from (16) and $\mathbf{E} = -\nabla\Phi = e\, r^{-2} \cdot \mathbf{e}_r$ from (10):

(20)   $f^r_{A\psi} = -\frac{\alpha}{4\pi}(f^2_{1s} + g^2_{1s})/r^2$

Combining the confinement force density (19) with the Lorentz force density (20) leads to the force density balance for the 1s state:

(21)   $f^r_\psi + f^r_{A\psi} = -\frac{1}{4\pi}[2 f_{1s} g_{1s}/r^2 + 2(f_{1s} g'_{1s} - f'_{1s} g_{1s})/r + (f_{1s} g''_{1s} - f''_{1s} g_{1s})] - \frac{\alpha}{4\pi}(f^2_{1s} + g^2_{1s})/r^2$

The evaluation of this expression is simplified by the fact that the 2nd and 3rd terms of $f^r_\psi$ vanish for the 1s state [6]. One then easily verifies explicity with the wave function (15) that the two force densities compensate each other for the 1s state.

It is informative to analyze the force density balance according to powers of the fine structure constant $\alpha$. At face value, the confinement force density seems to be of $O(\alpha^0)$ and the Lorentz force density of $O(\alpha^1)$. But both need to be multiplied by a factor of $O(\alpha^3)$, since the probability density $\psi^*\psi$ is normalized to a volume comparable to the cube of the Bohr radius $a_0 = \alpha^{-1} \cdot \hbar/m_e c$. And both acquire another factor of $O(\alpha^2)$ when evaluated at $r = a_0$. That still seems to leave the confinement force density short by a factor of $\alpha$. This factor can be found in the prefactor $(m_e - E_{1s})^{1/2}$ of $f_{1s}$ in (15). As a result, both force densities scale like $\alpha^6$ at $r = a_0$, and the force balance holds for any value of $\alpha$.

### 3. Conclusions and Outlook

This exercise has fulfilled its goal of testing explicitly whether a local force balance exists at every point inside the hydrogen atom. That is stronger than the commonly-used global stability criteria. The Coulomb force density is straightforward to calculate, but the opposing confinement force density requires evaluating the divergence of the stress tensor for the Dirac field. Approximating the proton in H by a point charge allows analytic expressions for the two opposing force densities.

A local force balance might seem odd for a quantum-mechanical system, where one would expect such detailed information not to be available due to the uncertainty relation. This is indeed true for a single experiment, which would not allow a complete measurement of the force density at all points. But it can be determined with arbitrary accuracy by repeating the same experiment often enough to obtain sufficient statistics. The force density behaves just like a probability density, which can be measured by repeated scattering experiments.

The force density balance is derived for a classical Dirac field, which corresponds to a single-particle wave function in quantum physics. It would be interesting to find out whether a local force balance holds for manybody systems in quantum field theory. For example, one could ask whether radiative corrections, such as the Lamb shift, maintain the local force balance in the H atom. A more fundamental manybody problem will be proposed below, i.e., the stability of the vacuum polarization surrounding a point charge.

Transferring the canonical formalism to manybody systems leads to a Lagrangian containing an infinite number of classical fields, each representing a real or virtual particle. A possible implementation would be the Hartree-Fock approximation of quantum electrodynamics developed by Greiner et al. (Ch. 16 in [5]). The key question arises whether or not the necessary renormalization procedures affect the force balance. Manybody states behave very differently from single-particle wave functions, as evidenced by P. W. Anderson's orthogonality catastrophe [7]. In that case an impurity introduced into an electron gas affects the manybody state dramatically, while the single-electron wave functions are only weakly perturbed.

Returning to the long-standing question about the internal stability of the electron [1],[2],[3], one has to realize that the traditional model of a charge cloud interacting with its own electric field becomes obsolete in a more realistic many-electron picture. An electron is not allowed to interact with itself anymore – at least not directly. Any many-electron wave function, such as a Slater determinant, is antisymmetric under the exchange of two electrons. A self-interacting electron appears twice in the determinant, once as the source of an electromagnetic field and a second time as the particle interacting with the field. Consequently the Slater determinant vanishes. In Hartree-Fock theory, the self-Coulomb interaction exactly cancels the self-exchange interaction. Thus, the classical self-interaction becomes an indirect process, where an electron first interacts with surrounding vacuum electrons and those subsequently interact with the original electron.

Such arguments suggest investigating the stability of a point charge embedded into the vacuum of quantum electrodynamics. At distances larger than the reduced Compton wavelength of the electron ($\lambdabar_C = \hbar/m_e c$) the vacuum may be viewed as a plasma consisting of virtual electrons and positrons. Virtual particle/antiparticle pairs with larger masses are suppressed exponentially. As a consequence, any charged particle behaves like a point charge. Other properties, such as mass, spin, and even the magnetic moment are irrelevant, since the magnetic dipole potential decays faster than the Coulomb potential. Thus, the point charge model applies to all charged particles at large distances.

The universal asymptotic Coulomb field polarizes virtual electrons and positrons in the vacuum of quantum electrodynamics, creating a charge cloud that reaches out to $\lambdabar_C$. Since the induced charges oppose the inducing point charge (see Ch. 14 in [5]), the situation is analogous to the H atom. The vacuum polarization charges are attracted to the external point charge, while their confinement within $\lambdabar_C$ leads to a repulsive force. In a many-electron system the exclusion principle causes an additional Pauli repulsion, similar to the degeneracy pressure which stabilizes white dwarf stars against gravitational collapse [8].

The Lorentz force density acting on the vacuum polarization is obtained in straightforward fashion from the corresponding charge density $\rho_{VP}$ given in the Appendix. A quantitative comparison with the Lorentz force density acting on the 1$s$ electron in the H atom is given in Figure 1. If one could get a handle on the repulsive force densities, one could again try to balance the force densities and analyze their dependence on $\alpha$. If they were all of the same order in $\alpha$, a force balance would hold for arbitrary values of $\alpha$ (as in the H atom). But if the forces scaled differently with $\alpha$, a balance could only be achieved at a specific value of $\alpha$. That would open the door to a theoretical determination of the fine structure constant.

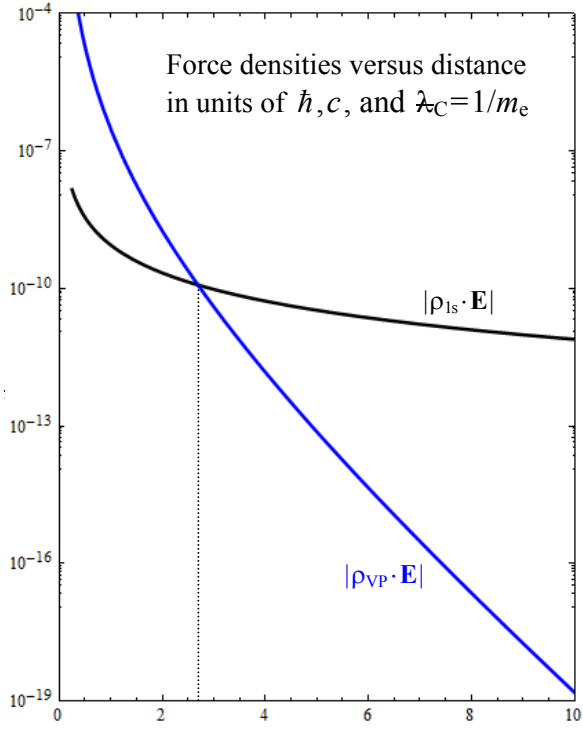

**Figure 1** The radial Lorentz force densities acting on the H 1s electron charge density $\rho_{1s}$ and the vacuum polarization charge density $\rho_{VP}$. The semi-logarithmic scale reveals two exponential decay lengths, half of the Bohr radius $a_0 \approx 5 \cdot 10^{-11}$m and half of the reduced Compton wavelength $\lambdabar_C \approx 4 \cdot 10^{-13}$m. $\lambdabar_C$ serves as length unit and $m_e$ as mass unit. At distances below the dotted line ($\approx 10^{-12}$m) the force balance is dominated by manybody effects. That is much larger than the proton radius of $\approx 10^{-15}$m, where nuclear forces take over. $\rho_{1s}$ is taken from (15),(16) and $\rho_{VP}$ from the Appendix.

The vacuum of quantum electrodynamics contains not only virtual electrons and positrons, but also virtual photons. These lead to the Casimir force which can be viewed as the radiation pressure exerted by confined photons [9],[10]. While their interaction with the vacuum polarization would be of higher order in $\alpha$, their interaction with the point charge itself could lead to an independent force balance. Casimir initially found an attractive force [9] which had the correct power law to counteract the internal Coulomb repulsion of a classical charge cloud. But the force was later calculated to be repulsive for a spherical geometry with reflective boundary conditions [10]. An analogous force is exerted by confined virtual electrons, and it was also found to be repulsive for a spherical geometry [10]. Such calculations of the Casimir force depend on the boundary conditions (reflective vs. absorbing), which is easily rationalized by the radiation pressure model. It is not obvious, however, which boundary conditions hold in the case of a charged elementary particle.

To summarize, calculating manybody force densities is a hard problem, but it could be highly rewarding. The force densities contributing to a force balance might scale differently with $\alpha$ and thus allow a theoretical determination of the long-sought fine structure constant. Such a concept was already proposed by Casimir for the radiation pressure of vacuum photons holding a classical charged particle together. In that case it seems not to have worked out. Could it work for the quantum-mechanical forces acting on the vacuum polarization surrounding a point charge? This topic is addressed in a separate publication [12].

**Appendix: Notation, Solutions of the Dirac Equation, Vacuum Polarization**

$\hbar, c$ are used as units, together with the Gaussian system for electromagnetism. The metric tensor $g^{\mu\nu}$ has the signature $(+---)$. Common four-vectors are: $A_\mu = (\Phi, -\mathbf{A})$, $x^\mu = (t, \mathbf{r})$, $\partial_\mu = \partial/\partial x^\mu = (\partial/\partial t, \nabla)$, $\partial^\mu = g^{\mu\nu}\partial_\nu = (\partial/\partial t, -\nabla)$, $p^\mu = (E, \mathbf{p}) \to i\partial^\mu = (i\partial/\partial t, -i\nabla)$. Four-vectors have Greek indices with automatic summation over equal indices, while three-vectors have Latin indices.

The standard 4×4 Dirac matrices $\gamma^\mu = (\gamma^0, \boldsymbol{\gamma})$ are used. They consist of 2×2 blocks containing either the 2×2 unit matrix (for $\gamma^0$) or the vector of Pauli matrices $(\tau^1, \tau^2, \tau^3) = \boldsymbol{\tau}$. For writing the Dirac equation in spherical coordinates it is useful to have the projections $(\gamma^r, \gamma^\theta, \gamma^\varphi)$ of the spatial Dirac matrices $\boldsymbol{\gamma}$ onto the unit vectors $\mathbf{e}_r, \mathbf{e}_\theta, \mathbf{e}_\varphi$. This is accomplished via the corresponding Pauli matrices $(\tau^r, \tau^\theta, \tau^\varphi)$:

$$\tau^r = \boldsymbol{\tau}\cdot\mathbf{e}_r = \begin{bmatrix} \cos\theta & \sin\theta\, e^{-i\varphi} \\ \sin\theta\, e^{i\varphi} & -\cos\theta \end{bmatrix} \quad \tau^\theta = \boldsymbol{\tau}\cdot\mathbf{e}_\theta = \begin{bmatrix} -\sin\theta & \cos\theta\, e^{-i\varphi} \\ \cos\theta\, e^{i\varphi} & \sin\theta \end{bmatrix} \quad \tau^\varphi = \boldsymbol{\tau}\cdot\mathbf{e}_\varphi = \begin{bmatrix} 0 & -i\, e^{-i\varphi} \\ i\, e^{i\varphi} & 0 \end{bmatrix}$$

The spherical Pauli spinors $\chi_\kappa^m(\theta,\varphi)$ are given by:

$$\chi_\kappa^m(\theta,\varphi) = \frac{1}{\sqrt{2}}\begin{bmatrix} -\text{sign}(\kappa)\cdot[1 - m/(\kappa+\tfrac{1}{2})]^{1/2} \cdot Y_l^{m-1/2}(\theta,\varphi) \\ [1 + m/(\kappa+\tfrac{1}{2})]^{1/2} \cdot Y_l^{m+1/2}(\theta,\varphi) \end{bmatrix} \qquad \int \chi_\kappa^{m\dagger}\chi_\kappa^m\, d\Omega = 1$$

with $\quad \kappa = \pm 1, \pm 2, \ldots \qquad \kappa = (-1)^{l+1}\cdot(j+\tfrac{1}{2}) \qquad l = |\kappa+\tfrac{1}{2}| - \tfrac{1}{2} \qquad j = |\kappa| - \tfrac{1}{2}$

The vacuum polarization induced by a point charge $+e$ can be expressed analytically in terms of the modified Bessel and Struve functions $K_n$ and $\mathbf{L}_n$:

$$\rho_{VP} = -e\,\frac{\alpha}{3\pi}\, m_e^3 \cdot \{1 - 2z\, K_2(2z)\cdot \mathbf{L}_{-3}(2z) - [4K_0(2z) + (4z^{-1}+2z)\cdot K_1(2z)]\cdot \mathbf{L}_{-2}(2z)\}$$

$$\rho_{VP} \to -e\,\frac{\alpha}{4\pi^{3/2}}\, m_e^3 \cdot z^{-5/2}\, e^{-2z} \qquad \text{for } r \to \infty \qquad\qquad z = m_e r$$

This expression is derived from the Laplacian of the electrostatic potential given in [11] for the leading order in $\alpha$. The following explicit expression is used for the integral function of $K_0$:

$$\int_z^\infty K_0(z')\, dz' = \{1 - z\cdot[K_0(z)\cdot \mathbf{L}_{-1}(z) + K_1(z)\cdot \mathbf{L}_0(z)]\}$$

**References**

1. The electron self-interaction problem is posed elegantly in the Richard Feynman's Lectures on Physics, Vol. II, Ch. 28, Caltech (1964) and in his Nobel lecture. For reviews see: S. Schweber, *An Introduction to Relativistic Quantum Field Theory*, Row, Peterson & Co., Evanston (1961), Section 15a; Ph. Pearle, *Classical Electron Models*, Ch. 7 in: *Electromagnetism, Paths to Research*, ed. by D. Teplitz, Plenum Press, New York (1982); F. Rohrlich, *Classical Charged Particles*, Addison-Wesley, Redwood City (1990).
2. For early work on the electron self-interaction see: H. Poincaré, Comptes Rendus **140**, 1505 (1905); H. A. Lorentz, *The Theory of Electrons*, 2nd Edition, Dover, New York 1952, Sections 26-37, 178-183 (this work goes back to lectures in 1906); E.